\documentstyle[11pt,newpasp,twoside]{article}
\def\gtwid{\mathrel{\raise.3ex\hbox{$>$\kern-.75em\lower1ex\hbox{$\sim$}}}}
\def\ltwid{\mathrel{\raise.3ex\hbox{$<$\kern-.75em\lower1ex\hbox{$\sim$}}}}

\def\edcomment#1{\iffalse\marginpar{\raggedright\sl#1\/}\else\relax\fi}
\reversemarginpar

\begin{document}
\title{Dubious Deductions from AGN Survey Data}

\author{Robert Antonucci}
\affil{University of California, Physics Department, Santa Barbara, CA 93106,
USA}

\begin{abstract}The participants in this meeting are almost all carrying out the
hard work of making many different types of AGN surveys.  Since it's so much
easier to criticize other people's work than to do actual work myself, I'll
just present some demurs regarding recent papers drawing conclusions from
various AGN survey data.  In particular I'll mention some questionable
interpretations of surveys of Seyfert 2 near-UV polarization; interpreting
the results of searches 
for polarized broad H-alpha lines in Seyfert 2s; testing the beam model for
radio galaxies and quasars; testing the unification of Seyfert spectral types
with a torus; and finally testing the energy sources for Ulirgs, especially
those with Liner optical spectra.  Only the polarized broad H-$\alpha$ results
are examined in detail here.
\end{abstract}

\footnotetext{The Proceedings of
the Colloquium will be published by the Astronomical
Society of the Pacific as a volume of the ASP Conference Series. The
editors will be Richard Green (Editor-in-Chief, USA), Edward Khachikian
(Armenia) and David Sanders (USA).}

\section{Levels of polarization of the nuclear light in Seyfert 2 galaxies}

Koski (1978) showed that Seyfert 2s and Narrow Line Radio Galaxies generally
have UV excesses relative to the SEDs for the Pop II stars which dominate the
optical light in arcsec apertures.  He showed the spectra could be
parameterized approximately by
a normal Pop II SED, plus a suitably normalized power law which dominates in
the near-UV.  The brightest powerful Seyfert 2 nucleus, that of NGC1068, can 
indeed be fit quite well that way (Miller and Antonucci 1983, McLean et al
1983, Antonucci et al.\ 1994), and furthermore the power-law component has a
wavelength-independent polarization of $\sim16$\% in arcsec apertures.  It
is crucial to note that the broad permitted
lines show the same percent P as the power-law continuum component accounting
for the ``UV excess."  This strongly suggests that the scattering paths of
the broad
lines and the power law are very similar geometrically.  Finally, the permitted
lines have normal equivalent width in polarized flux.  This confirms that the
hidden source is simply a normal Seyfert 1 nucleus, whose light is scattered
into the line of sight.

The polarization angles of NGC1068 and other Seyfert 2s lie perpendicular to the
radio axes, so that the photons' last flights before scattering into the line
of sight were {\it along}\/ the axes.  Thus the equatorial directions must be
opaque,
and the torus model was born.  In principle the polar scattering could result
from intrinsically anisotropic emission, but the anisotropy would need to be
the same for the broad emission lines as for the power law, which would be
difficult to arrange by some mechanism other than shadowing.  In fact if these
two components differ substantially in their isotropy in any significant number
of objects, we'd expect to see one component without much of the other fairly
frequently in Type 1 objects.  The broad line equivalent width dispersion
would then be very large and this is not seen.

  It was evident from the beginning of the era of high SNR spectropolarimetry 
that most other objects behave somewhat differently (Antonucci 1984 and Tran
et al.\ 1995 on 3C234;  Miller and Goodrich 1990;  Tran 1995).
It's quite general that the broad line equivalent widths
are normal in the polarized flux spectra of narrow line objects, that is, they
match those of broad line objects.  Thus the basic picture is the same as for
NGC1068:  a normal broad line nucleus is seen in reflection.  But in general
the UV ``featureless continuum" has a polarization much lower than that of the
broad emission lines (Kay 1994).  Thus there is another, unpolarized UV
continuum component, now know to derive from hot stars in many cases (Heckman
et al.\ 1995, Gonzalez-Delgado et al.\ 1998).  But the key from the unified model
point of view is that the broad lines are highly polarized (at 16\%, NGC1068
has about the {\it lowest}\/ P), so that the scattering is in one or two
polar cones.

  The point here is that several authors have taken the percent P from the UV
continuum---which is often low---and rejected polar scattering on that basis.
For example, see the key role of this erroneous argument in the discussion
of Malkan et al.\ 1998.\footnote{The way to measure the broad line polarization,
and hence the polarization of the scattered light alone, is to divide the
polarized flux by the total flux.  However, it is generally impossible to see
the line clearly in total flux, so one just derives high lower limits to the broad
line polarization in most cases.  Tran (1995) claimed to measure the broad lines in
total flux in some objects.  However, M.~Kishimoto and I looked carefully
at the case of Mrk477, in preparing a paper on polarization imaging.  We
examined the total-flux plots, overlaying the permitted Ba lines on various
forbidden lines; we saw little or no evidence for broad wings in the total flux.
We thus placed an upper limit on the total flux of the broad
components, leading to a conservative lower limit on broad H-$\alpha$ and
H-$\beta$ polarization of 10\%.  Thus we disagree with the intrinsic
value of only 2--3\% quoted in Tran 1995.  We used the same data.}

\section{Polarized broad H-$\alpha$ and the generality of Seyfert unification}

Since the early days of the geometrical unified models of various kinds, attempts
have been made to assess the generality of the models.  For example, Tran
(2001) has recently discussed whether or not all Seyfert 2s have hidden
Type 1 nuclei of any significant luminosity.  It's usually difficult to
prove that something is {\it not present}\/, and people haven't yet converged
on the answer to this.\footnote{One way to do it is to see whether there's
any significant ``waste heat"---dust reradiation from the matter obscuring
the putative hidden nucleus.  D.~Whysong and I are studying 3C radio galaxies
in the mid-IR with this in mind, and have shown, for example, that M87 can
have no such hidden nucleus at any remotely relevant luminosity level:
astro-ph 0106381.  Also, Meisenheimer et al.\ (2001) present a very good study
of ISO data, and reach a similar conclusion that many lower-luminosity
radio galaxies lack significant waste heat.}

Tran (2001) has given a variety of arguments and concludes that there are
in fact many ``real" Seyfert 2 galaxies without hidden Type 1 nuclei.  His
paper may be correct but I want to take this opportunity to review some
places where his arguments seem quite uncertain---most of these are already
discussed in his paper.

Tran (2001) presents spectropolarimetric data on two Seyfert 2 samples:  the
$12\mu$-selected sample of Rush et al.\ 1993, and the CfA sample of Huchra and
Burg 1992.
The main finding seems to be that those with detected hidden broad emission
line regions (``HBLRs") have higher radio power relative to the far-IR power
than the non-HBLRs on average, and that they also have warmer
$25\mu$--$60\mu$ colors on average.  These differences are taken to show that
the two types (HBLRs vs.\ non-HBLRs\footnote{Note that ``non-HBLR'' in Seyfert
2s means
there is no BLR present at all, since the Seyfert 2 classification precludes a
visible BLR.}) of Seyfert 2
are intrinsically different, and that in particular there is a large subset
with no hidden Type 1 nucleus.

I've often preached about the great improvement in interpretability of tests
of this type when the samples are selected by a property which is thought
to be fairly isotropic, such as far-IR, and H.~Schmitt explores this in detail
in his article in this volume.  (See also Schmitt et al.\ 2001.)  That's
certainly not the case with the $12\mu$
sample, since the very high column densities of the tori result in substantial
optical depths at that wavelength, though this sample is much better in this
regard than those based on say, UV excess.  Aside from all theoretical
considerations, and aside from reference to the X-ray columns, it seems
obvious that the $12\mu$ emission is quite anisotropic simply because the HBLR
Seyfert 2s have much steeper mid-IR spectra than bare Type 1 nuclei (see e.g.,
Edelson \& Malkan 1986).  This
wouldn't be the case if the mid-IR were emitted isotropically.  The X-ray
columns reinforce this:  they are {\it much}\/ larger in the Type 2s of all
kinds, than in the Type 1s, and this huge extra column density is probably
dusty molecular gas.  That last point is confirmed semi-quantitatively at
least from molecular line maps.\footnote{The X-ray columns in Type 1 nuclei,
while much smaller in general, are still usually larger than those expected
from the optical reddening/extinction estimates, for Galactic dust size
distributions and dust/gas ratios  (Maiolino et al.\ 2001 and references
therein).  It's possible that {\it this}\/ gas contains only large grains,
or perhaps is dust-free, but it's quite a different issue from the enormous
excess columns of Type 2s relative to Type 1s.}  See for example Planesas
et al.\ 1991.  Thus I can't agree with the
characterization of the $12\mu$ sample as complete and
unbiased with respect to orientation.

For a few important related fine points of Tran's paper,
he asserts that the contradictory conclusions of Heisler et al.\ 1997 may be due
to sample ``incompleteness" of the latter.  As discussed below and in Schmitt's
contribution to these proceedings, I believe the selection of the latter
is in fact better, based largely on the far-IR which is thought to be
relatively isotropic.  The Heisler paper was recently updated as Lumsden
et al.\ 2001, and the interpretation given there is more consistent with
Tran's interpretation.

Other caveats about Tran's paper:

1)  The polarimetry data come from Lick, Palomar and Keck Observatories.  
No indications are made that the observations are of uniform depth, and
no upper limits are given for broad H-alpha for the putative non-HBLRs.
It would not be simple to rectify this situation.  However, I already see
from a footnote to Moran et al.\ 2001
that {\it two of Tran's ``non-HBLR'' objects do show the broad H-$\alpha$
in polarized flux}\/ with better data.

2)  The warm mid-IR average colors of the HBLR objects have been discussed
before (e.g. Heisler et al.\ 1997,  Lumsden et al.\ 2001;  Alexander 2001, and
references therein;  see also Miller and Goodrich 1990), with arguments as
to whether the warmer colors might derive from lower inclinations as expected
for tori theoretically.  That could explain the lack of polarized broad
H$\alpha$ if the objects
with the highest inclinations have partially obscured scattering regions.
There is also evidence given in these
papers that at least part of the effect (the correlation between detectability
of a scattered BLR and mid-IR ``warmth'') is a contrast issue related to the
relative dominance of the AGN in a particular object, and this seems natural
and unavoidable too. 

Tran argues (following Alexander 2001) that the cooler mid-IR spectra of
the non-HBLRs cannot be due to higher extinction to the warm regions (and thus
presumably to torus inclination) because the X-ray columns for the samples
under consideration are statistically
indistinguishable.  To me that's suggestive at best because of the sample
mismatches, and more importantly, because it's a purely qualitative
argument.  No theory predicts the magnitude of the expected difference
between the columns of the HBLRs and non-HBLRs if the difference is due to
inclination.  One would need to have such a robust prediction to compare
with any observational {\it upper limit}\/ on the statistical column density
difference, in order to prove a discrepancy.

3)  A point that I find at least strongly suggestive (given
the problematic selection criteria), is that the HBLR Seyfert 2s have a larger
ratio of radio/far-IR than the others.\footnote{The paper also cites Moran
et al.\ 1992 as providing evidence for intrinsically greater absolute radio 
luminosities in HBLR Seyfert 2s;  that result was subject to severe selection
effects, and was essentially retracted in Moran et al.\ 2000.
Certainly given the anisotropic selection criteria for the Tran objects,
I'd hesitate to make much of the radio power difference in the Tran paper.}
This seems to be a modest effect though (see Tran's Fig.~1; no statistical
significance is given in the paper, but a significance of 0.8\% is given by
a K-S test: H.T., pc).  
If it is accepted to be statistically significant for the two populations,
a next step would be to examine whether free-free absorption in the radio
would be significant for the objects of highest inclination.  Many recent
papers on Seyfert radio properties have concluded free-free optical
depths can be substantial in the cm region (e.g., Gallimore et al.\ 1997, 
Ulvestad 1999, and several others).
Also a starburst component might have a different radio/far-IR ratio.
But a different starburst contribution doesn't constitute strong evidence 
against a hidden BLR.

Tran and also Thean et al.\ 2001 find that the absolute radio luminosities of
the HBLR Seyfert 2s are significantly greater than those of the non-HLBRs
for the 12$\mu$ sample.  However, with imperfect selection criteria, I'm
more comfortable with comparing ratios of nearly isotropic properties such as
discussed in the previous paragraph.

Also a general caveat should be kept in mind.
The fact that a statistical difference in the populations of the putative types
of 2 may exist is only suggestive of a qualitative difference in the physics.
If all have hidden BLRs, the two types would necessarily differ in average
torus covering factor, so needn't be identical in other intrinsic properties.

4)  The point is made that the [OIII] 5007/ H-$\beta$ ratio median is
$9.9\pm1.3$
for the HBLR nuclei but only $6.8\pm1.5$ for the non-HBLRs.  This brings up
the vexing question of definitions for Seyfert 2.  Well known examples
of bright Seyfert 2s have ratios near 10.  Half of the ``non-HBLR Seyfert
2s" as used in the paper have ratios less that 6.8.  The ratios aren't
tabulated in the paper, but I'd interpret this as meaning that many so-called
non-HBLRs are composites, with important Liner or starburst contributions
to many of them.  Certainly these components will reduce the integrated dust
temperatures in the sense observed.  But again that doesn't constitute
evidence against a hidden BLR commensurate with the amount of high-ionization
narrow line gas present.

5)  Tran asserts that the non-HBLR Seyfert 2s simply have no hidden BLR.
But the BLR is always accompanied by a ``power-law" continuum, and this
continuum is the only viable explanation for some of the narrow line ratios.
Thus the question arises:  how are the non-HBLR narrow line regions 
ionized?  He speculates that the latter are ``dominated by other 
nuclear and circumnuclear processes such as starbursts."  This seems 
consistent with the statement that they have lower excitation, but also 
seems to reduce the question to semantics.  No one doubts that there
are objects dominated by e.g. starbursts, which thus have different narrow
line ratios than Seyferts\dots it's just that the starbursts are relatively
more important in some objects than in others, and the excitation is
a measure of that.  Again I don't see any implication that there's no
hidden BLR commensurate with the requirements for narrow line ionization
level in a particular object.\footnote{Nuclear activity seems manifest
mainly as the Big Blue Bump and the resulting broad and narrow emission
lines in radio quiet AGN.  There is no question
that some radio galaxies lack significant visible {\em or} hidden BBBs,
based on the mid-IR argument mentioned in Footnote 2, but they are still
AGN because of the radio activity.  But what AGN activity
exists in a non-HBLR non-radio-loud Seyfert 2?
In finance, the analogy would be a bond which is a ``zero coupon perpetuity,''
which makes no interest payments, and never pays back the principle.  At
that point it's not much of a bond, and a non-HBLR Seyfert 2 might be
similarly ill-defined.}

6)  The paper of Pappa et al.\ (2001) is cited as showing the existence
of low intrinsic X-ray absorption in two Seyfert 2s of very low luminosity.
Many Seyfert 2's are Compton thick and thus show no absorption turnover
at low X-ray energies, but for these two, other arguments are given that
suggest that this isn't likely.  As Pappa et al.\ point out, the lack of
an observed BLR could be intrinsic, or else it could be due to a dusty
warm absorber as documented in several more luminous objects.  I'm not aware 
that either of these two, NGC3147 or NGC 4698, has been checked for broad
polarized H-$\alpha$, but that would certainly be worth while.

My personal conclusion from all this is that the existence of objects
without the Big Blue Bump and accompanying broad emission lines---commensurate
with the luminosity of highly excited narrow line gas---has not been shown
robustly.  This is actually pretty similar to Tran's
conclusion that ``it is the strength of the AGN engine that seems to be the 
dominant factor in determining the visibility of the HBLR."
Undoubtedly if the nuclear continuum and BLR strength can be turned down in
a particular object, everything else being the same, the scattered light signal
would be harder to see.  The question is, does anything happen
{\it qualitatively}\/ at low AGN luminosities?  Do these components decrease
faster than say the
strength of the high-excitation narrow line gas?  It seems to me we're as
far as ever from answering that question.

\section{Testing the generality of Seyfert unification with isotropic
properties}

It would be great to have a dollar for every paper that concluded that Seyfert
1s and 2s are intrinsically different by showing that samples differ in some
way, without selecting the sample by an isotropic property.  No unified model
says that any old batch of objects of one type is equivalent to any old batch
of objects of another type.  As an extreme example, with UV selection, an
object whose UV excess is just scattered light must be much more powerful
intrinsically than one whose UV excess is seen directly.  For objects like
NGC1068, only a percent or so of the nuclear UV is scattered into the line
of sight, so Seyfert 2s found in this way come from five magnitudes higher
on the luminosity function than Seyfert 1s found this way.  No wonder they
have more CO, L(far-IR), radio emission, etc, etc!  I'll just mention one
recent example.  The Malkan et al 1998 paper, based on the Seyferts in the
HST archives, concludes that Seyfert 2s are more likely to have nuclear
dust structures.  According to the verbal contribution
of Maiolino et al.\ to the Gullermo Haro Workshop held at UNAM in 2000,
and in a pc, this effect disappears if the Seyfert types are matched for [OIII]
luminosity.  (Other demurs regarding Malkan et al.\ 1998 can be found in
Antonucci 1999a.)

A must-read paper in this context is Keel et al.\ 1994, which analyzes
properties of Seyferts selected by $60\mu$ flux and $25\mu$--$60\mu$ color.
That sample should be pretty good---maybe the best that is available right
now.  Several old saws about Seyfert 2s having greater narrow line
luminosities, ratios of narrow line luminosities to radio, etc, are
disproven there.  Not everything in Keel et al.\ can be easily
explained by orientation though!

H.~Schmitt is leading a major program of study of the sample from the Keel
et al.\ paper, and he gives a report in these proceedings.  (See also Schmitt
et al.\ 2001.)  He shows that several
claimed differences between Seyfert 1s and 2s go away with this type of
selection.  To make the sample even better, we hope to add in the Seyferts
dropped from the Keel et al list because of the $25\mu$--$60\mu$ color
criterion.  The result should be quite good.

There is a crucial limitation of this type of test, however.  As
emphasized in the past by e.g., A.~Lawrence, there {\it must}\/ be a range of
covering factors for the dusty tori, and so those classified as Type 2
{\it must}\/ have higher average covering factors.  Therefore as populations,
the 1s and 2s {\it must}\/ be intrinsically different in their statistical
properties at some level.  This has {\it no direct implication}\/ that
some objects lack hidden BLRs!

To reiterate, even a perfect study, using isotropic selection and real
upper limits, is expected to show differences between 1s and 2s in their
intrinsic statistical properties.  One could only confidently expect broad 
overlap in the properties of the 1s vs.\ the 2s even if every 2 has a hidden
Type 1 nucleus.  Such a difference would by itself have no direct
implication that some of the Type 2s lack a BLR. 

\section{Testing the beam model for radio galaxies and quasars with isotropic
properties}

This is an old issue, pretty much settled long ago I think.  It still carries
a good lesson, very closely analogous to that of the previous section.
Blandford and Rees (1978) and Blandford and Konigl (1979) proposed that
superluminal radio sources were simply normal double sources seen from
very low inclinations, i.e., seen from along the axes of the radio jets.  Early
tests of the idea using the associated radio projected linear sizes
(Browne\footnote{Browne was, however, an early advocate for matching diffuse
radio power.}
et al.\ 1982) and the narrow emission lines (Heckman 1983) found major
discrepancies with it.  However the double-lobed objects used in the
comparison samples to the superluminals came from orders of magnitude higher
on the luminosity function, because they satisfied survey flux limits
with their nearly isotropic diffuse radio emission, whereas the superluminals
satisfied survey flux limits because of their beamed core emission.  It was
exactly like the Seyfert 1/2 sample problems referred to in the last section.
This was discussed in detail in Antonucci and Ulvestad 1985.

\section{Testing the energy sources for ultraluminous infrared galaxies with
mid-IR spectroscopy}

As for the previous section, I'll give just a brief sketch of this
important issue, with a reference to a recent fuller discussion: 
Antonucci 2001. 

The issue is the energy sources in Ultraluminous Infrared Galaxies, and whether
recent ISO mid-IR spectral surveys provide breakthroughs in this area.
  
A large subset of the Ulirgs have properties {\it exactly}\/ as
expected for ultraluminous ``Quasar 2s," that is to say powerful
high-ionization narrow line emission, powerful mid-far infrared emission,
no optical point sources, but instead diffuse ``mirror" regions revealing
light from a hidden quasar.  Note that the optical continuum is not expected
to scale with nuclear luminosity because in low luminosity Seyfert 2s, it's
almost always dominated by the underlying old stellar population.

The narrow emission lines have much lower equivalents widths in quasars than
in Seyfert 1s, so their luminosity doesn't scale proportionately to the nuclear
optical/UV either (e.g., Boroson \& Green 1992, and Wills et al.\ 1993).
It's a big effect.

As with nearby Seyfert 2s, it's
difficult to be very precise about any starburst contribution to Ulirg
energetics, but some well-studied cases reveal that it is often substantial.
However, it is almost universally true that the objects characterized by
Seyfert-2 like narrow line ratios in the optical show the same type of
emission line spectrum in the mid-IR.
Thus, the latter spectral region just reveals the same emission component as
the optical for this group of Ulirgs.

At lower luminosities especially, many Ulirgs have starburst optical spectra,
and also starburst mid-IR spectra.  Again the mid-IR observations reveal
just the same emission region as the optical, at least in a qualitative
sense.

Where the mid-IR data show something new is among the many Ulirgs with 
{\it LINER}\/ optical spectra:  in almost all of these cases the mid-IR spectra
show lines from HII regions (e.g., Lutz et al.\ 1999)!  This is interpreted
most simply (and in the published papers, essentially solely) as proving that
the true nuclei, and by implication the dominant energy sources, are compact
starbursts.  However, in general the star formation
uncovered this way cannot be shown to dominate (or otherwise) the galaxy
energetics because there is no accurate conversion from any particular
starburst spectral feature to the bolometric luminosity of the associated
starburst population.

Thus the ISO data show the presence of substantial but poorly determined
starburst luminosity somewhere in the galaxies, generally {\it at the
obscuration depth penetrated by light of these wavelengths}\/.  Unfortunately,
obscured AGN are known to possess column densities of
$\gtwid10^{24}$ cm $^{-2}$ in most cases, where $A(V)\gtwid1000$ and the
mid-IR from the nucleus is quite obscured.  There is a moderate range of
parameter space such that the mid-IR can't penetrate the gas/dust columns,
but the hard X-rays can.  Above $>10^{25}$ cm $^{-2}$ or so however,
the dusty gas becomes ``Compton thick" and even the hard X-rays are blocked.

The point is that recently many Ulirgs classified as LINERS in the optical
and starbursts in the mid-IR
have since revealed powerful $\sim10$keV X-ray sources, often with luminosities
suggestive of hidden AGN which are capable of providing the entire
observed bolometric power.  In other LINER Ulirgs a hidden powerful AGN is
revealed by spectropolarimetry, though with that technique we have
no good way of estimating the contribution of the AGN to the bolometric
luminosity.  These cases show that published conclusions that $>50$\% of the
energy in a Ulirg derives from a hidden starburst are invalid.  The same
applies to the Scuba sources, which were placed directly on the
``Madau" diagram of stellar luminosity density evolution, without the
slightest evidence that they're powered by stars!

The details and references are given in the review article cited at the top of 
this section (as well as in papers by experts such as Sanders and Mirabel,
and Veilleux and several others).  I just mention two cases here because
they are really amusing to me.  The well-studied Ulirg NGC6240 is a
{\it LINER}\/ in the optical, but a starburst in the mid-IR.  In fact it's
a ``template"
starburst in the mid-IR according to Genzel et al.\ 1998.  The Genzel team
claims to have shown that this and many other examples are starburst-
dominated on this basis.  Only trouble is, NGC6240 and many others
have powerful AGN X-ray sources coming directly through the obscuring
matter at 10keV (Vignati et al.\ 1999).

I don't know of any starburst spectral feature that can be accurately
used to determine the starburst bolometric luminosity, so the starburst
features can only say that there is such a component, and put a relatively
low minimum value on its luminosity.  The situation is a little better
for AGN in that the X-ray luminosity is empirically a pretty good
($\sim$~factor of 3?) estimator of the bolometric value, based on the
unobscured cases.  Vignati et al.\ (1999) point out that that relation
indicates that the bolometric luminosity of the hidden AGN in NGC6240 is
consistent with the entire observed bolometric luminosity.  Conservatively,
I conclude from that that the AGN makes a significant contribution, though
it's not certain that it's $>50$\%.

The situation for NGC4945 is similar.  Spoon et al.\ (2000) have analyzed the
mid-IR spectrum finding that the starburst ``may well power the entire
bolometric luminosity, \dots[but] are also consistent with an up to 50\%
contribution from an embedded AGN."  Immediately afterward, Madejski et
al.\ (2000)
published a spectacular wide-band SED showing the few times $10^{24}$ X-ray
column, and deriving the unabsorbed nuclear X-ray luminosity alone at
$\sim10^{43}$ erg/s, a substantial fraction of that in the mid-far-IR,
and indicative of a bolometric luminosity consistent with that observed.
It is very important to note that there are no indications of activity in
NGC4945 other than the hard X-ray source.  This means that active galaxies can
look perfectly normal at other wavelengths!  They do not need to show even
the liner-like emission lines!

\section*{Acknowledgements}

For comments on an earlier draft I thank L.~Kay, T.~Heckman, B.~Wills, 
M.~Kishimoto, H.~Schmitt, J.~Ulvestad, and R.~Barvainis.  Detailed comments
from B.~Wills led to many changes.  Support came
from NSF grants NSF AST96-17160 and NSF AST00-98719.

\end{document}